\documentclass[]{extarticle}
\usepackage{graphicx}
\usepackage{natbib}
\usepackage{times}

\bibpunct{(}{)}{;}{a}{}{}
\usepackage{mathtext,amssymb}

\newcommand{\Mach}{\ensuremath{\mathcal{M}}}

\begin{document}

\title{On the Maximal Value of the Turbulent
  $\alpha$-Parameter in Accretion Discs}
\author{P. Abolmasov\thanks{Corresponding author: {\it pasha@sao.ru}}~ and N. I. Shakura\\
{\it \small Sternberg Astronomical Institute, Moscow State University,
  Moscow, Russia 119992}}



\maketitle

\abstract{In this short paper we show that making turbulence two- rather than
three-dimensional may increase effective turbulent viscosity by about 40\%.
Dimensionless hydrodynamical viscosity parameter up to $\alpha_{max} = 0.25 \Mach_t^2$ may be
obtained in this approach, that is in better agreement with the
observational data on non-stationary accretion than the values
obtained in numerical simulations. However, the $\alpha$-parameter
values known from observations are still several times higher. 
}

\maketitle

\section{Introduction}

Though $\alpha$ prescription \citep{shakura72,SS73} proved to be quite useful for
accretion disc physics, the physical mechanisms driving angular
momentum transport are still not completely understood. Hydrodynamic as
well as MHD turbulence and magnetic fields are generally
proposed as the main viscosity sources. One of the sources of MHD
turbulence is magneto-rotational instability first predicted in the
classic works by \citet{velikhov} and \citet{chandra60} and applied
for the case of disc accretion by \citet{BH91}.

Stationary accretion disc properties depend weakly on the
viscosity $\alpha$-parameter itself. $T_{eff} \propto R^{-3/4}$ effective temperature
profile (and hence the outcoming multicolor blackbody spectrum, see \citet{lyndenbell69}) is
independent of viscosity.
However, most of the physical quantities (such as density,
radial velocity and the total mass of the disc)
depend somehow on the viscosity parameter.

Detailed spectral fitting is capable for viscosity parameter estimates through
the density of the disc atmosphere. Few attempts (such as \citet{davis06}) were made yet to
determine $\alpha$-parameter from the X-ray spectra,
but these estimates are still uncertain and highly model-dependent. 

Due to that reason, $\alpha$-parameter is generally determined from
observational data on non-stationary accretion phenomena such as dwarf
nova flares and X-ray transients
\citep{Smak99,King07,Sul08}. The actual viscosity parameter 
value generally affects
characteristic viscosity timescales and subsequently the shapes of
optical and X-ray lightcurves. 
\citet{qiao09}
use transitions between the low-hard and high-soft states in X-ray
binaries to estimate the actual $\alpha$-parameter values in the range $0.1\div 0.9$.

The dimensionless viscosity parameter is proven to vary in the range
$0.1\div 0.9$ for hot, fully ionized accretion discs. Numerical
MHD simulations generally predict $\alpha \lesssim 0.02$ (see
\citet{King07} for review) much smaller than the values required by observations.

Slightly higher values such as $0.1\div 0.2$ appear in hydrodynamical turbulence models
such as that of \citet{Kato94}. Turbulence is treated as a
three-dimensional phenomenon. There are however
reasons to consider also the two-dimensional case because vertical turbulent
motions are damped by positive vertical entropy gradient \citep{SSZ78,tayler80}.
Here we follow the general line of
\citet{Kato94,nara94,KY97,discs98}, but consider two-dimensional case instead
of three-dimensional. In the following section we write out the
general form of the Reynolds stress evolution equations for the 2D-case. In section
\ref{sec:stac} a steady-state axisymmetric solution is established. We estimate the
dimensional viscosity parameter value in section \ref{sec:alpha} and
discuss our results and make conclusions in sections \ref{sec:disc}
and \ref{sec:conc}.

\section{Reynolds Stress}\label{sec:stress}

Here we define the turbulent stress tensor as $t_{ij} =
\langle u_i u_j\rangle$, where $u_{i,j}$ are the fluctuating velocity components. The
flow is considered incompressible, that presumably does not affect a
non-divergent flow.
We modify the equations by \citet{KY97} in a way to account for the different
dimensionality of the case considered. In our case, anisotropy
tensor is naturally defined as $b_{ij} = (t_{ij} - K)/2K$, 
where $i,j = r, \varphi$ and $K =
(t_{rr} + t_{\varphi\varphi})/2$. Dissipation rate $\varepsilon$
should also appear without the $2/3$ miltiplier present in the
three-dimensional approach. Stress tensor evolution equations may
be written as follows:

\begin{equation}\label{E:init:trr}
\left( \frac{\partial}{\partial t} + \Omega \frac{\partial}{\partial
  \varphi} \right) t_{rr} =  4\Omega t_{r\varphi} + \Pi_{rr} - \varepsilon
\end{equation}

\begin{equation}\label{E:init:tphiphi}
\left( \frac{\partial}{\partial t} + \Omega \frac{\partial}{\partial
  \varphi} \right) t_{\varphi\varphi} =  -\frac{\varkappa^2}{\Omega} t_{r\varphi} + \Pi_{\varphi\varphi} - \varepsilon
\end{equation}

\begin{equation}\label{E:init:trphi}
\left( \frac{\partial}{\partial t} + \Omega \frac{\partial}{\partial
  \varphi} \right) t_{r\varphi} =  2\Omega t_{\varphi\varphi} -
\frac{\varkappa^2}{2\Omega} t_{rr} + \Pi_{r\varphi}
\end{equation}

For the pressure-strain correlation terms $\Pi_{ij}$ we adopt the general form
with two dimensionless constants, used by \citet{Kato94}, 
$\Pi_{ij} = -C_1 K \varkappa b_{ij} + C_2 K S_{ij}$. $C_{1,2}$ are
universal dimensionless constants, and $S_{ij} = (\partial U_i /
\partial x_j + \partial U_j / \partial x_i) /2$ is the rate of strain
of the avaraged flow.
 Following \citet{Kato94}, we use
$\varkappa$ rather than $\Omega$ \citep{discs98} as the characteristic frequency for
local processes such as inertial waves.
That assumption makes no difference for the case of Keplerian rotation
when $\varkappa = \Omega$. Results may be modified by substituting
$C_1 \Omega / \varkappa$ instead of $C_1$ in all the equations. 
We ingore the two additional terms with coefficients $C_3$ and
$C_4$ \citep{KY97}. Their inclusion (for any $C_4 > 0$ and $0< C_3 < 2$) may only
decrease the $t_{r\varphi}$ stress component value.

Finally, pressure-strain correlation terms take the form:

\begin{equation}\label{E:Prr}
\Pi_{rr} = -\frac{C_1}{2} \varkappa\left(t_{rr} - K\right)
\end{equation}

\begin{equation}\label{E:Pphiphi}
\Pi_{\varphi\varphi} = -\frac{C_1}{2} \varkappa\left(t_{\varphi\varphi} - K\right)
\end{equation}

\begin{equation}\label{E:Prphi}
\Pi_{\varphi\varphi} = -\frac{C_1}{2} \varkappa t_{r\varphi} +
\frac{C_2 }{2} r \frac{d\Omega}{dr} t_{r\varphi}
\end{equation}


\section{Steady-State Axisymmetric Solution}\label{sec:stac}

Stationary solution is easily obtained from equations
(\ref{E:init:trr}-\ref{E:init:trphi}) by zeroing their left-hand
sides. 
It is instructive also to obtain an expression for the steady-case
dissipation term as a sum of equations (\ref{E:init:trr}) and
(\ref{E:init:tphiphi}). Kinetic energy $K$ is also conserved in
stationary case, hence:

\begin{equation}
0 = \frac{{\rm d}K}{{\rm d}t} = -r\frac{{\rm d}\Omega}{{\rm d}r} t_{r\varphi} -\varepsilon
\end{equation}

The dissipation rate
may be therefore written as $\varepsilon = -r d\Omega / dr
t_{r\varphi}$. Note that the two pressure-strain correlation terms
exactly zero after summation.
After some algebra one obtains the components of the Reynolds stress
tensor (normalised by $2K$) as follows:

\begin{equation}\label{E:stasol:trr}
\begin{array}{l}
\frac{t_{rr}}{2K} = \frac{1}{2} \left( 1 - \frac{2}{A_0} \left( 1 -
\frac{C_2}{2}\right) \times  \right. \\
\qquad{}\qquad{}\qquad{} \qquad{} \left. \times \left(2 + \frac{1}{2}\frac{d\ln \Omega}{d\ln r}
\right) \frac{d\ln \Omega}{d\ln r}\right)
\end{array}
\end{equation}

\begin{equation}\label{E:stasol:tphiphi}
\begin{array}{l}
\frac{t_{\varphi\varphi}}{2K} = \frac{1}{2} \left( 1 - \frac{1}{A_0} \left( 1 -
\frac{C_2}{2}\right)  \times \right. \\
\qquad{}\qquad{}\qquad{} \qquad{} \left. \times \left(-\frac{\varkappa^2}{\Omega^2} + \frac{d\ln \Omega}{d\ln r}
\right) \frac{d\ln \Omega}{d\ln r}\right)
\end{array}
\end{equation}

\begin{equation}\label{E:stasol:trphi}
\frac{t_{r\varphi}}{2K} = - \frac{C_1}{4 A_0}
\frac{\varkappa}{\Omega} \left( 1-\frac{C_2}{2}\right) \frac{d\ln \Omega}{d\ln r},
\end{equation}
\noindent

where:

\begin{equation}\label{E:stasol:A0}
A_0 = \left( \frac{C_1^2}{4} + 4\right) \frac{\varkappa^2}{\Omega^2} +
\left( \frac{d\ln \Omega}{d\ln r}\right)^2
\end{equation}


\section{Estimating Turbulent $\alpha$-Parameter}\label{sec:alpha}

Let us estimate the turbulent viscosity parameter for the
two-dimensional case. We define the dimensionless $\alpha$ parameter as:

$$
\alpha = \frac{\rho |t_{r\varphi}|}{p} = \frac{\gamma |t_{r\varphi}|}{c_s^2},
$$
\noindent
where $c_s$ is the speed of sound, and $\Mach_t^2 = 2K / c_s^2$ 
is therefore the turbulent Mach number squared. For the general case of
arbitrary $C_{1,2}$:

\begin{equation}\label{E:alpha:gen}
\alpha = \frac{\gamma C_1}{4 A_0} \frac{\varkappa}{\Omega} \left( 1-\frac{C_2}{2}\right) \left|
\frac{d \ln \Omega}{d \ln r} \right| \Mach^2_t
\end{equation}

As in the three-dimensional case, $\alpha$-parameter has a maximum at a
certain value of $(C_1)_{max}$. In the two-dimensional case
$(C_1)_{max} =5$ for Keplerian rotation. The maximal value of $\alpha$
is:

\begin{equation}\label{E:alpha:max}
\alpha_{max} = \frac{\gamma}{8} \frac{\varkappa}{\Omega}
\frac{1}{\sqrt{1+\frac{1}{4} \frac{\Omega^2}{\varkappa^2} \left(
    \frac{d \ln \Omega}{d \ln r}\right)^2}} \left|
\frac{d \ln \Omega}{d \ln r} \right| \Mach^2_t
\end{equation}

Maximal $\alpha$-parameter values for two- and three-dimensional case
differ by about 40\%. In the particular case of Keplerian rotation, $C_2 = 0$ and
$\gamma=5/3$,
\begin{equation}\label{E:alpha:2D3D}
\alpha_{max} = \left\{ \begin{array}{lc}
0.18 \Mach^2_t & \mbox{for 3D}\\
0.25 \Mach^2_t & \mbox{for 2D}\\
\end{array}\right.
\end{equation}

The second expression is the precise upper limit for two-dimensional
hydrodynamic turbulence. The difference with the three-dimensional
case is that all the turbulent energy is concentrated in radial and
azimuthal motions that contribute to the radial angular momentum
transfer. Presence of the additional $z$ dimension leads to decrease
of radial and azimuthal motion amplitudes. 

\section{Discussion}\label{sec:disc}



The viscous $\alpha$-parameter may be defined in different ways.
Above we used $\alpha = \alpha_p = \rho t_{r\varphi} /
p$. Alternatively, $\alpha$ parameter is sometimes defined using the 
turbulent viscosity coefficient $\nu = \alpha_\nu c_s^2/\Omega$. 
These two definitions of the $\alpha$-parameter are related by the
following expression:

\begin{equation}\label{E:alpha:SS}
\alpha_p = \frac{\rho t_{r\varphi}}{p} = \frac{\gamma \nu}{c_s^2}
r\left|\frac{d\Omega}{dr} \right| = \gamma \left| \frac{d\ln \Omega}{d\ln r} \right| \ \alpha_\nu 
\end{equation}

The two parameter definitions $\alpha_p$ and $\alpha_\nu$ are
absolutely equivalent.
For Keplerian rotation and $\gamma = 5/3$, $\alpha_p$ is higher by a
factor $2.5$.
Therfore, our $(\alpha_p)_{max}=0.25$ corresponds to $(\alpha_\nu)_{max}=0.1$.


Vertical structure of a thin accretion disc is unfavourable for strong
turbulence development in z-direction. If net entropy increases with
the distance from the equatorial plane of the disc, vertical motions are
expected to be damped. That is the case for
standard accretion discs in gas pressure-dominated regime \citep{SSZ78,tayler80}, excluding
the zones of partial ionization of hydrogen where super-adiabatic
temperature gradient appears \citep{meyer82}.
Stable vertical stratification affects mostly the largest
turbulent spatial and velocity scales, where vertical motions are effectively damped by
buoyancy forces. 
 The situation is different for horizontal turbulent
motions that always have an energy influx from the
averaged shear flow. Thin accretion discs are predicted to be unstable
with respect to radial convective motions (Rossby wave instability,
see \citet{MH90,lovelace99,slava07}).
Largest vortices of the turbulent cascade receive their energy
and momentum immediately from the averaged flow. Due to that
they are expected to be close to complanar 
with the accretion disc rotation. On the other hand, vertical motions
do not interact directly with the mean
flow. Because the mean flow is primarily affected by the largest
eddies of the turbulent cascade,
 turbulent velocity field is effectively two- rather
than three-dimensional, when its dynamical effect on the mean flow is
considered.


Excitation mechanisms and energy balance of hydrodynamical turbulent motions in
accretion discs are tightly connected with the instabilities present
in the flow.  Though thin accretion discs generally fulfil the
 Reynolds and H{\o}iland stability criteria, they may be unstable to
 non-axisymmetric perturbations in the presence of radial entropy
 gradient \citep{lovelace99}. Besides this, even in the linearly stable
case there are certain ways to destabilize differentially rotating
flows through non-linear instabilities, mode coupling and transient
growth solutions (see \citet{UR04} for review and references therein).
Recurrent transient growth solutions may lead to large turbulent
energy growth for the large Reynolds numbers characteristic for
accretion discs \citep{Afshordi05}.
In short, it seems that turbulence is likely to develop even in
linearly stable cases. Simulations are not consistent however in this
conclusion -- certain numerical simulations \citep{Hawley99} report
stabilization of all the instability modes in shearing box
simulations.


Turbulence excitation in differentially rotating fluids are being
extensively studied in laboratory experiments with Taylor-Couette
flows.
 A comprehensive explanation of the first experiments of that kind \citep{Taylor36} was given by \citet{Zeldovich81}, who also points out
the similarity between rotating fluid motions and astrophysical
discs. 
So far, laboratory experiments do not support high viscosity values in
differentially rotating fluids with quasi-Keplerian
profiles. Effective $\alpha$ values found by \citet{Ji2006} are well  below
$\sim 10^{-5}$ for high Reynolds numbers ${ \rm Re }\gtrsim 10^6$.
 Taylor-Couette experiment, however, differs from astrophysical
discs in some important points like radial entropy gradient and
gravitational field of the central body. Resulting non-axisymmetric
convective instability modes may be therefore responsible for high
accretion rates in astrophysical discs. Besides this, astrophysical
discs are highly supersonic flows, while Mach numbers in laboratory
experiments are much lower unity. 

Numerical simulations mostly confirm the existence of long-living turbulence
in rotating shear flows. Large scale two-dimensional 
vortices, mainly of anticyclonic
nature, are reported to form and 
survive \citep{Bodo07,Lithwick09,Shen06}, sometimes
exceeding disc thickness in their spatial extent in azimuthal
direction. \citet{Bodo07} report that the azimuthal size of a typical vortex
exceeds the scaleheight of the disc for low speed of sound values 
(or high Mach numbers of the flow), starting from about 0.05 in Keplerian velocity units. 
So the presence of the large-scale two-dimensional vortices
relevant for angular momentum transfer is well confirmed by numerical
simulations. It is not clear whether the vortices themselves are
stable if the effect of the third (vertical) direction is taken into
account. \citet{Lithwick09} claims that vortex dynamics and stability
properties do not change significantly in 3D with respect to the 2D
case. On the other hand, another recent paper by \citet{LP09} report
that elliptic 2D vortices are almost always destroyed by resonance
processes with the local epicyclic oscillations. 



\section{Conclusions}\label{sec:conc}

Our main conclusion is that the turbulent viscosity $\alpha$-parameter
may increase by up to 40\% if the turbulence is two-
rather than three-dimensional. The resulting upper limit $(\alpha_p)_{max}
= 0.25 \Mach_t^2$ is still too
low to explain the high viscosity parameter values derived from
non-stationary accretion phenomena. 

It is possible that the contemporary turbulence theory based either on
laboratory experiments or a quasi-particle approach
misses some important effects leading to more effective
turbulent motion interaction.
Alternatively, one may allow turbulence to be supersonic or to
consider the influence of magnetic fields. The impact of the latter in
controversial: as it was already mentioned in Introduction, MHD
simulations predict relatively low $\alpha$-parameter values.

We conclude that the observational viscosity parameter is still
2$\div$3 times higher than the upper limits given by theory. 
The problem is yet to be solved.

\bigskip
This work was supported by the RFBR grant 09-02-00032. We thank the
referee (Marek Abramowicz) for useful comments and suggestions.

\bibliographystyle{mn2e}
\bibliography{mybib}

\label{lastpage}

\end{document}